\documentclass[aip,apl,a4,reprint,twoside,floatfix,superscriptaddress]{revtex4-1}

\usepackage{graphicx}
\usepackage{amsmath}
\usepackage{amssymb}
\usepackage{amsfonts}
\usepackage{dcolumn}
\usepackage{epsfig}
\usepackage{subfigure}
\usepackage{bm}

\begin{document}

\title{$|\epsilon|$-Near-Zero materials in the near-infrared}

\author{Alessandro Ciattoni}
\email{alessandro.ciattoni@aquila.infn.it} \affiliation{Consiglio Nazionale delle Ricerche, CNR-SPIN, 67100 Coppito L'Aquila, Italy}

\author{Rino Marinelli}
\affiliation{Dipartimento di Ingegneria Elettrica e dell'Informazione, Universit\`{a} dell'Aquila, Via Giovanni Gronchi 18 - Zona industriale di
Pile, 67100 L'Aquila,Italy}

\author{Carlo Rizza}
\affiliation{Consiglio Nazionale delle Ricerche, CNR-SPIN, 67100 Coppito L'Aquila, Italy} \affiliation{Dipartimento di Fisica e Matematica,
Universit\`{a} dell'Insubria, Via Valleggio 11, 22100 Como, Italy}

\author{Elia Palange}
\affiliation{Dipartimento di Ingegneria Elettrica e dell'Informazione, Universit\`{a} dell'Aquila, Via Giovanni Gronchi 18 - Zona industriale di
Pile, 67100 L'Aquila,Italy}

\begin{abstract}
We consider a mixture of metal coated quantum dots dispersed in a polymer matrix and, using a modified version of the standard Maxwell-Garnett mixing
rule, we prove that the mixture parameters (particles radius, quantum dots gain, etc.) can be chosen so that the effective medium permittivity has an
absolute value very close to zero in the near-infrared, i.e. $|\textrm{Re}(\epsilon)| \ll 1$ and $|\textrm{Im}(\epsilon)| \ll 1$ at the same
near-infrared wavelength. Resorting to full-wave simulations, we investigate the accuracy of the effective medium predictions and we relate their
discrepancy with rigorous numerical results to the fact that $|\epsilon| \ll 1$ is a critical requirement. We show that a simple method for reducing
this discrepancy, and hence for achieving a prescribed value of $|\epsilon|$, consists in a subsequent fine-tuning of the nanoparticles volume
filling fraction.
\end{abstract}
\maketitle

Epsilon-near-zero (ENZ) materials have attracted much attention in the last decade for their intriguing optical properties \cite{Silveirinha}. The
 $|\epsilon|$-near-zero condition $|\epsilon| \ll 1$ (where both real and imaginary parts of the dielectric permittivity are very small) have
suggested novel ways to achieve remarkable effects like transmissivity directional hysteresis \cite{Ciatt_1}, peculiar plasmonic memory
functionalities and bistability \cite{Ciatt_2} and highly efficient second-harmonic generation \cite{Ciatt_3}. All of these intriguing effects are
consequences of the fact that, in the presence of the very small permittivity background, nonlinearity is no longer a perturbation thus disclosing a
novel and highly nonlinear optical behavior. The scenario is additionally supplied by a novel kind of field enhancement consisting in the fact that,
in the transverse magnetic configuration, the longitudinal component of the electric field becomes singular at the interface between vacuum and the
$|\epsilon|$-near-zero medium, due to the continuity of the displacement field component normal to the interface \cite{Scalora}. So it is quite
natural to ask: \textit{to what extent can we realize a $|\epsilon|$-near-zero material at the prescribed optical wavelength?}

It is well-known that standard ENZ materials with the less restrictive condition $\textrm{Re}{(\epsilon)} \simeq 0$ are directly available in nature.
Main examples are low-loss metals (like Au and Ag) whose Drude-type dispersion behavior is such that, close to their ultra-violet plasma frequency,
the real part of the dielectric permittivity is very small \cite{Cai}. In addition heavily doped semiconductors as ITO, GZO and AZO have Drude-type
dispersion behavior as well (transparent conductors) and they can be doped to tune their plasma frequency to point of achieving the condition
$Re{(\epsilon)} \simeq 0$ at telecom wavelengths \cite{Gordon,Hiramatsu}. On the other hand there are semiconductors exhibiting Lorentz-type
dispersion behavior as, for example, GaP, GaAs, and Si for which the condition $\textrm{Re}{(\epsilon)} \simeq 0$ occurs in the ultra-violet
\cite{Palik}. In the visible range natural ENZ materials are lacking, so that the engineering of metal-based composite materials is required for
filling this gap \cite{Karri,Roberts,Rizza}. In the context of metamaterials, the condition $\textrm{Re}{(\epsilon)} \simeq 0$ can be achieved at a
prescribed wavelength by suitably designing the composite which has to comprise both positive and negative permittivities. In addition, if among the
metamaterial constituents there is a gain medium, even the imaginary part of the effective permittivity can be managed \cite{Klar,Wuestner,Xiao} and
this is a basic ingredient for achieving the $|\epsilon|$-near-zero condition. An efficient gain mechanism in the near infrared is provided by
quantum dots and they have been exploited by several authors to design lossless metamaterials \cite{Zeng,Fu}. Recently, Rizza et al. have shown that
a gain-assisted metallo-dielectric nano-laminate can be designed to fulfil the $|\epsilon|$-near-zero condition at the visible frequency
\cite{Rizza}, the loss compensation being achieved by incorporating optically pumped dye molecules (with a resonance wavelength of $\lambda=610$ nm)
in the dielectric layers.
%
%

\begin{figure}
\center
\includegraphics*[width=0.5\textwidth]{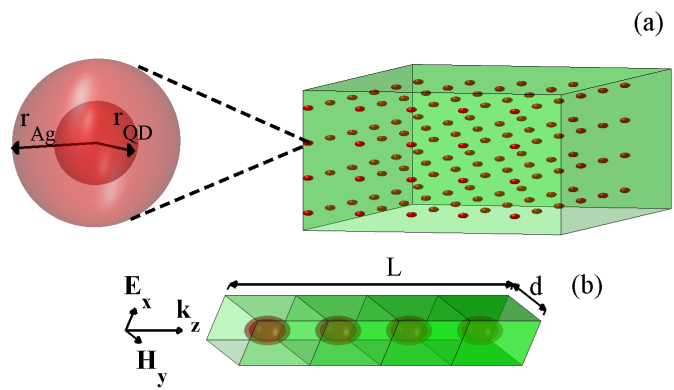}
\caption{(Color online) (a) Sketch of the considered composite material obtained by dispersing silver-coated PbSe/ZnSe quantum dots within a polymer
(PMMA) matrix. Here $r_{Ag}$ and $r_{QD}$ are the outer and inner radii of the nanoparticle silver shell, respectively. (b) Sketch of the impinging
field geometry and of the integration domain used for the full-wave analysis.}
\end{figure}

In this Letter we show that the condition $|\epsilon|$-near-zero can be achieved in the near-infrared through a composite material obtained by
dispersing silver-coated PbSe/ZnSe quantum dots within a polymer (PMMA) matrix (see Fig.1(a)). Exploiting a modified version of the Maxwell-Garnett
mixing rule which additionally accounts for the geometry of the structured nanoparticles \cite{Zeng}, we tailor the mixture properties (e.g. the core
and shell radii and the quantum dot gain) so that both the real and imaginary of the effective permittivity are very close to zero at the same
near-infrared wavelength. It is worth noting that $|\epsilon| \ll 1$ is a very critical condition, its actual fulfilling being an highly nontrivial
task. In fact in an actual manufactured sample, a variety of generally minor effects (departure from homogeneous behavior, spatial nonlocal effects,
surface contributions, etc.) can produce a large relative deviation of the dielectric permittivity from the very small value predicted by the
effective medium theory. Therefore we have performed rigorous full-wave simulations to investigate the actual feasibility of the
$|\epsilon|$-near-zero condition in a slab filled by the proposed mixture. It turns out that the dielectric permittivity predicted by full wave
simulations is very small and of the some order of magnitude of that predicted by the effective medium approach, the relative discrepancy being
however significant. On the other hand we show that such a discrepancy can be reduced by means of a further designing stage which consists in
slightly adjusting the nanoparticle volume filling fraction.

Within the context of the effective medium approach, if the radiation wavelength is much greater that the nanoparticle radius, we can regard the
silver-coated spherical PbSe/ZnSe quantum dots as an effective homogeneous sphere whose equivalent permittivity is \cite{Zeng}
\begin{equation} \label{str}
\epsilon_{s}=\epsilon_{Ag} \frac{\epsilon_{QD}(1+2\rho)+2\epsilon_{Ag}(1-\rho)}{\epsilon_{QD}(1-\rho)+2\epsilon_{Ag}(2+\rho)},
\end{equation}
where $\rho=r_{QD}^3/r_{Ag}^3$ is the fraction of the total particle volume occupied by the inner material sphere ($r_{QD}$ and $r_{Ag}$ are the
inner and the outer shell radii, respectively, see Fig.1(a)), whereas $\epsilon_{Ag}$ and $\epsilon_{QD}$ are the PbSe/ZnSe quantum dot core and
silver shell permittivities, respectively. Here, the silver and the PbSe/ZnSe quantum dots dispersion behaviors are described by the Drude and
Lorentz-Drude model, respectively, i.e.
\begin{eqnarray}
\epsilon_{Ag}&=&\epsilon_{\infty}-\frac{\omega_p^2}{\omega^2+i\omega\Gamma}, \nonumber\\
\epsilon_{QD}&=&\epsilon_{b}+A \frac{\omega_0^2}{\omega^2-\omega_0^2+i2\omega\gamma},
\end{eqnarray}
where $\epsilon_{\infty}=4.56$, $\omega_p=13.8 \cdot 10^{15}$ Hz, $\Gamma=0.3\cdot 10^{15}$ Hz, $\epsilon_b=12.8$, $\omega_0=2.27\cdot 10^{15}$ Hz,
$\gamma=1.51 \cdot 10^{12}$ Hz \cite{Holmstrom,Fu} and $A$ is a dimensionless parameter related to the gain efficiency of the considered quantum dot.
Note that the quantum dots resonant wavelength is $\lambda_0=(2 \pi c)/\omega_0 = 827$ nm so that the mixture is adequate for the near-infrared.
Exploiting the standard Maxwell-Garnett approach, the effective dielectric permittivity of the polymer (PMMA) matrix hosting the dispersed metal
coated quantum dots is
\begin{equation} \label{eff}
\epsilon=\epsilon_{h} \frac{\epsilon_{s}(1+2f)+2\epsilon_{h}(1-f)}{\epsilon_{s}(1-f)+2\epsilon_{h}(2+f)},
\end{equation}
where $f$ is the nanoparticle volume fraction and $\epsilon_{h}=2.202$ is the PMMA permittivity. For the geometrical parameters $r_{QD}=5$ nm,
$r_{Ag}=6.23$ nm ($\rho=0.52$) and the volume filling fraction $f=0.1$, the effective medium approach of Eqs.(\ref{str}) and (\ref{eff}) yields
$\epsilon=0.01$ ($\textrm{Im}(\epsilon)=0$) at $\lambda = 827$ nm for $A=0.003$. In Fig.2, we report the real (panel (a)) and imaginary (panel (b))
parts of the effective dielectric permittivity $\epsilon$ of Eq.(\ref{eff}) (solid line), as a function of $\lambda$ in a spectral range around the
quantum dots resonant wavelength $\lambda_0$. We conclude that the proposed mixture can actually be tailored to fulfill the $|\epsilon|$-near zero
condition.

\begin{figure}
\center
\includegraphics*[width=0.5\textwidth]{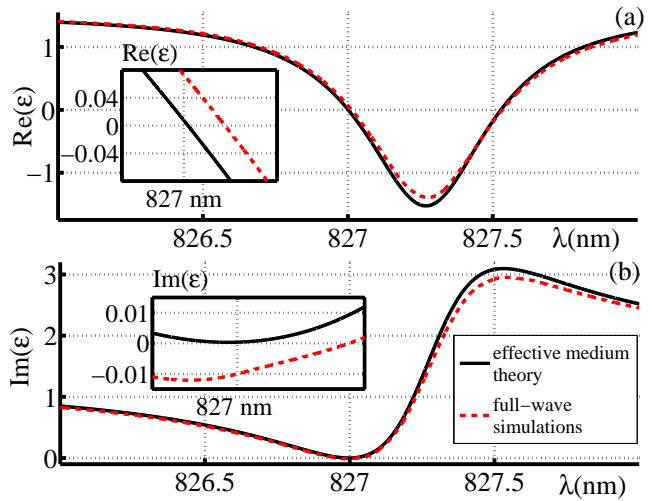}
\caption{(Color online) Real (panel (a)) and imaginary (panel (b)) parts of the composite dielectric permittivities evaluated through the effective
medium approach of Eq.(\ref{eff}) (solid lines) and retrieved from the full-wave simulations (dashed lines) as functions of the wavelength $\lambda$.
The insets of panels (a) and (b) are magnified copies of the corresponding panels for wavelengths close to $\lambda=827$ nm.}
\end{figure}

In most of the proposed configurations \cite{Ciatt_1,Ciatt_2}, the $|\epsilon|$-near-zero condition is not exploited for representing the limiting
theoretical situation where the permittivity strictly vanishes ($\epsilon=0$) but the actual numerical small value of $|\epsilon|$ plays a
fundamental role since it tunes various effects and field enhancements \cite{Ciatt_3,Scalora}. It is evident that manufacturing an actual sample
exhibiting the prescribed small value of $|\epsilon|$ can be a very nontrivial task since a variety of generally minor effects (departure from
homogeneous behavior, spatial nonlocal effects, surface contributions, etc.) can produce a large relative deviation of the dielectric permittivity
from the very small value predicted by the effective medium theory. In order to investigate the actual feasibility of the proposed
$|\epsilon|$-near-zero mixture, we have performed 3D finite-element full-wave simulations \cite{Comsol} for evaluating the transmission coefficient
of a slab filled with the proposed mixture and consequently retrieving its effective dielectric permittivity \cite{Smith}. More precisely, we have
considered a parallelepiped of height $L=86.49$ nm of equal width and thickness $d=21.6$ nm (see Fig.1(b)), containing four metal-coated quantum dots
aligned along the $z$-axis and we have adopted periodic boundary conditions along the surfaces parallel to the $z$-axis. Two layers of unit
permittivity (vacuum) have been located at the top and bottom parallelepiped faces for providing normal incidence illumination with monochromatic
light and for collecting the transmitted field.
\begin{figure}
\center
\includegraphics*[width=0.5\textwidth]{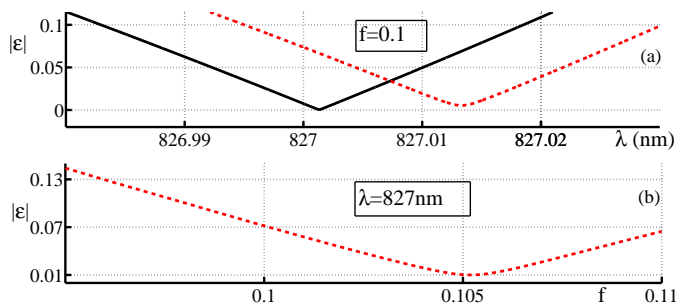}
\caption{(Color online) (a) Moduli of the theoretical effective dielectric permittivity evaluated from Eq.(\ref{eff}) (solid line) and retrieved from
full-wave simulations (dashed line) as functions of the wavelength $\lambda$. (b) Modulus of the dielectric permittivity retrieved from full-wave
simulations as a function of the nanoparticle volume filling fraction $f$.}
\end{figure}
It is worth noting that we have set the geometrical sizes of such a numerical integration domain to coincide with those employed in the above
described effective medium analysis. In Fig.2, in addition to the effective medium approach results (solid lines), we have plotted the real and
imaginary parts of the dielectric permittivity retrieved from full wave simulations (dashed lines) and we note that there is a very good global
agreement between theoretical predictions and numerical results. On the other hand it is evident from the insets of panels (a) and (b) of Fig.2, that
in the spectral range around the $\textrm{Re}(\epsilon)=0$ crossing point, the \textit{relative} discrepancy $\left(\Delta |\epsilon|\right) /
|\epsilon|$ is significant since, for example, at $\lambda=827$ nm, full-wave analysis yields $|\epsilon|=0.07$ (see panel (a) of Fig.3) that is
seven times larger than the expected value $|\epsilon| = 0.01$. Such a discrepancy is unavoidable it resulting from all the physical effects and
mechanisms neglected by the effective medium theory. However, we here point out that in the design of the actual medium, such a discrepancy can be
essentially reduced by means of a second-stage parameter choosing which consists in refining the nanoparticle volume filling fraction. In Fig.3(b) we
plot the $|\epsilon|$ retrieved from further full-wave simulations that we have all performed at the wavelength $\lambda=827$ nm for different
nanoparticle volume filling fraction $f$. We note that the requirement $|\epsilon| = 0.01$ is attained for $f=0.105$, i.e. a slight change of $f$
allows the actual mixture to exhibit the required value of $|\epsilon|$ at the specified wavelength.

In conclusion, we have exploited the optical properties of a mixture of metal coated quantum dots dispersed in a polymer matrix to prove that, by
suitably choosing its geometrical parameters, it is possible to achieve a very small value of the absolute value of the permittivity at an infrared
wavelength. We believe that the realization of a material exhibiting the $|\epsilon|$-near-zero property in the near-infrared or in the visible
ranges is a very important target in the field of photonics since, the very small permittivity would allow each possible permittivity change (which
is generally very small) to play a fundamental role on electromagnetic propagation. As a consequence, $|\epsilon|$-near-zero materials are expected
to offer very original ways for achieving efficient light steering through external control (via optical nonlinearity, electro-optic effect or
acousto-optic effect, etc.) and therefore for designing a future generation of photonic and nano-hotonic devices.

The research leading to these results has received funding from the Italian Ministry of Research (MIUR) through the 'Futuro in Ricerca' FIRB-
grant PHOCOS - RBFR08E7VA.


\begin{thebibliography}{1}
\bibitem{Silveirinha} M. G. Silveirinha and N. Engheta, Phys. Rev. Lett. {\bf 97}, 157403 (2006).
\bibitem{Ciatt_1} A. Ciattoni, C. Rizza and E. Palange, Opt. Lett. {\bf 35}, 2130 (2010).
\bibitem{Ciatt_2} A. Ciattoni, C. Rizza, and E. Palange, Phys. Rev. A {\bf 83}, 043813 (2011).
\bibitem{Ciatt_3} A. Ciattoni, http://arxiv.org/abs/1103.2864 (2011).
\bibitem{Scalora} M. A. Vincenti, D. de Ceglia, A. Ciattoni, M. Scalora, http://arxiv.org/abs/1107.2354 (2011).
\bibitem{Cai} W. Cai and V. Shalaev, Optical Metamaterials: Fundamentals and Applications (Springer, Dordrecht, 2010).
\bibitem{Palik} E. D. Palik, Handbook of Optical Constants of Solids (Academic Press, London-New York (1985).
\bibitem{Gordon} R. G. Gordon, MRS Bulletin {\bf 25}, 52 (2000).
\bibitem{Hiramatsu} M. Hiramatsu, K. Imaeda, N. Horio and M. Nawata, Jour. Vac. Sci. Technol. A {\bf 16}, 669 (1998).
\bibitem{Karri} J. Karri and A. R. Mickelson, Silver Dielectric Stack With Near- Zero Epsilon at a Visible Wavelength,
                2009 IEEE Nanotechnol- ogy Materials and Devices Conference June 2-5, 2009, Traverse City, Michigan, USA.
\bibitem{Roberts} M. J. Roberts, S. Feng, M. Moran, and L. Johnson, J. Nanophoton. {\bf 4}, 043511 (2010).
\bibitem{Rizza} C. Rizza, A. Di Falco, A. Ciattoni, http://arxiv.org/abs/1105.5533 (2011).
\bibitem{Klar} T. A. Klar, A. V. Kildishev, V. P. Drachev and V. M. Shalaev, IEEE J. Sel. Top. Quantum Electron. {\bf 12}, 1106 (2006).
\bibitem{Wuestner} S. Wuestner, A. Pusch, K. L. Tsakmakidis, J. M. Hamm and O. Hess, Phys. Rev. Lett. {\bf 105}, 127401 (2010).
\bibitem{Xiao} S. Xiao, V. P. Drachev, A. V. Kildishev, X. Ni, U. K. Chettiar, H. K. Yuan and V. M. Shalaev, Nature {\bf 466}, 735 (2010).
\bibitem{Zeng} Y. Zeng, Q. Wu, D. H. Werner, Opt. Lett. {\bf 35}, 1431 (2010).
\bibitem{Holmstrom} P. Holmstr\"{o}m, L. Thylen, A. Bratkovsky, Appl. Phys. Lett. {\bf 97}, 073110 (2010).
\bibitem{Fu} Y. Fu, L. Thylen, H. Agren, Nano Letters, {\bf 8}, 1551 (2008).
\bibitem{Comsol} COMSOL, www.comsol.com.
\bibitem{Smith} D. R. Smith, S. Schultz, P. Markos, and C. M. Soukoulis, PRB {\bf 65}, 195104 (2002).
\end{thebibliography}
\end{document}